# Water-Planets in the Habitable Zone: Atmospheric Chemistry, Observable Features, and the case of Kepler-62*e* and -62*f*

L. Kaltenegger[1,2]* , D. Sasselov[2], S. Rugheimer[2]
[1]Max Planck Institute of Astronomy, Koenigstuhl 17, 69115 Heidelberg, Germany, kaltenegger@mpia.de
[2]Harvard-Smithsonian Center for Astrophysics, Cambridge, MA, 02138, USA

**Abstract:** Planets composed of large quantities of water that reside in the habitable zone are expected to have distinct geophysics and geochemistry of their surfaces and atmospheres. We explore these properties motivated by two key questions: whether such planets could provide habitable conditions and whether they exhibit discernable spectral features that distinguish a water-planet from a rocky Earth-like planet. We show that the recently discovered planets Kepler-62*e* and -62*f* are the first viable candidates for habitable zone water-planet. We use these planets as test cases for discussing those differences in detail. We generate atmospheric spectral models and find that potentially habitable water-planets show a distinctive spectral fingerprint in transit depending on their position in the habitable zone.

Subject headings: Astrobiology - atmospheric effects - methods: data analysis – Earth - planets and satellites: general – stars: individual (Kepler)

## 1. Introduction

The possible existence of Earth and super-Earths[1]-size planets covered completely by a water envelope has long fascinated scientists and the general public alike (Kuchner 2003, Leger et al.2004, Selsis et al 2007). No such planets are known in the Solar System but small bodies like Pluto are composed of substantial quantities of water though none are in the HZ. Ocean-planets that form outside the ice line and migrate inwards to the Habitable Zone (HZ) and beyond were defined in detail by Selsis at al. (2007) in that broader sense and are now known to exist thanks to mean density measurements of a few transiting exoplanets (see e.g. Gautier et al.2012, Cochran et al.2011, Gilliland et al.2013).

Until recently all known candidates for ocean-planets (Borucki et al.2013) were found orbiting very close to their stars. Such planets, e.g., Kepler-18*b*,-20*b*,-68*b*, are very hot due to the high stellar flux, which ensures a smooth transition from an interior water envelope to a steam atmosphere with no liquid surface ocean (Rogers&Seager2010; Valencia et al.2009). The discovery of many planetary systems with tightly packed inner planets by the *Kepler* mission has opened the prospect for getting mean densities of Earth-size planets in the HZ by transit-timing variations (TTVs) where radial velocity amplitudes are too small to measure (Lissauer et al.2011). Given the very low mean densities measured so far among the majority of such planets, e.g. those found in Kepler-11 (Lissauer et al.2013), Kepler-20 (Gautier et al.2012), Kepler-36 (Carter et al.2012), we anticipate that the first HZ super-Earths of radius below 2 Earth radii ($R_E$) are more likely to be water-rich planets than rocky silicate-rich ones. The recent discovery of the multiple transiting system Kepler-62, with two planets in its HZ (Borucki et al.2013), illustrates this point. We are motivated by the discovery of Kepler-62*e* and -62*f* to consider a more general approach to computing surface and atmospheric conditions on water-planets in the HZ, that could also form in situ with a high water content (Raymond et al 2007). Here we will refer to these planets as water-planets, given our assumption of theoretical interior models with pure water composition in an envelope surrounding a silicate- and metal-rich core, but no implicit assumption about liquid versus icy surface, or planet formation origin of the water. We focus on the interface between mantle and atmosphere with view of computing observable spectra.

---

[1] Super-Earth size is used here for planets with radii between 1.25 $R_\oplus$ and 2.0 $R_\oplus$.

Water-planets of Earth- to super-Earth sizes in the HZ fall into at least 2 types of interior geophysical properties, in terms of effect on their atmosphere: In a first Type, hereafter Type1, the core-mantle boundary connects silicates with high-pressure phases of water (e.g. Ice VI,VII), i.e. the liquid ocean has an icy bottom. In contrast in Type2 the liquid ocean has a rocky bottom, though no silicates emerge above the ocean at any time. The second type is essentially a rocky planet when viewed in terms of bulk composition. Both subtypes could possess a liquid ocean outer surface, a steam atmosphere, or a full cover of surface Ice I, depending on their orbit within the HZ and the magnitude of their greenhouse effect. Frozen water-planets should show a subsequent increase in the surface albedo value due to high reflective ice covering a frozen surface and will be most easily detected by direct imaging missions operating in the visible, while IR imaging search will preferential detect warm water-planets. Surface and atmospheric conditions on water-planets in the HZ have not been the subject of detailed studies. So far the work has focused on issues of evaporation (Kuchner 2003; Valencia et al.2009; Murray-Clay et al.2010) and boundary conditions to interior models (Valencia et al.2007, Grasset et al.2009, Fu et al.2010, Rogers&Seager 2010).

In this paper we develop an initial model for water-planet atmospheres to allow assessment of their observables with future telescopes. This model will be based on many explicit and implicit assumptions using Earth as a template but taking into account anticipated differences in outgassing, geochemical cycling, global circulation, etc. to assess their observables with future telescopes. In this paper we describe a set of atmospheric models and their underlying assumptions in section 2, how they could be applied to interpret Kepler-62*e* and -62*f* as water-planets in sections 3 and 4, and finish in section 5 with discussion and conclusions.

**2. Atmospheric Models for Water-planets – Basic Assumptions**

We assume here that the initial composition of icy planetesimals that assemble into water-planets is similar to that of comets, mostly $H_2O$, some $NH_3$ and $CO_2$. An initial composition of ice similar to that of comets leads to an atmospheric model composition of 90%$H_2O$, 5%$NH_3$ and 5%$CO_2$ (see also Leger et al.2004, who simplified the atmospheric photochemistry by assuming no CO nor $CH_4$). $NH_3$ is UV sensitive, photodissociates and is converted into $N_2$ and $H_2$ in a very short time frame, with $H_2$ being lost to space (see e.g. Leger et al.2003,Lammer et al.2007). This shows that water-planet atmospheres should have the same chemical constituents as ocean-land planets.

The carbonate-silicate cycle that regulates $CO_2$ on our own planet is effective due to weathering of exposed solid rock surface. Recent work (Abbot et al.2012) has shown that for an Earth-like planet, the carbonate-silicate cycle could continue to function largely unchanged for a continent surface fraction as low as 10%, with mid-ocean ridges taking over some of the recycling processes but argued that $CO_2$ cannot build up in the atmosphere of a Type2 water-planet without continents. Therefore the HZ for water-planets was inferred to be narrower than for planets with continents.

Note that the arguments presented in that work are generally only applicable to Type2 water-planets, i.e. those with very low water-mass fraction. For Type1 water-planets, i.e. Super-Earths with water-mass fractions like Earth ($2-5 \times 10^{-4}$) or above, the deep oceans are separated from the rocky interior via a layer of high-pressure ices. Therefore an alternative mechanism to modulate abundant gases like $CO_2$ and $CH_4$ will be required in water-planets. Such a mechanism would depend on the properties of their clathration in water over a range of very high pressures. Methane clathrates are water molecule lattices that trap $CH_4$ molecules as guests by virtue of multiple hydrogen-bonding frameworks. Under very high pressures (above 0.6GPa), common to super-Earth-size water-planets, methane clathrates undergo a phase transition in their structure to a form known as filled ice. Levi et al (2013) studied this transition and concluded that $CH_4$ would be transported efficiently through the high-density water-ice mantle of water-planets in filled ice clathrates and eventually released in the atmosphere. It is well known that $CO_2$ substitutes $CH_4$ readily in normal-pressure clathrates (see e.g. Park et al.2006, Nago&Nieto2011). We expect that this property can be extrapolated to filled ice, though this needs to be established. If so, this substitution will enable effective cycling of $CO_2$ through the atmosphere and oceans in water-planets residing in the HZ (e.g., with water oceans at the surface) and thus provide an atmospheric concentration feedback mechanism for $CO_2$ on water-planets. Abundant gases like $CH_4$ and $CO_2$ will be transported through the high-pressure phases of the planets mantle (filled ice and Structure I clathrates), through the liquid water ocean, then into the atmosphere, balanced by solubility in the ocean (depending on $CO_2$ solubility vs temperature), and the sequestration of $CO_2$ in clathrates. Similar to Earth-like planets with continents, such a cycle would control the $CO_2$ levels on water worlds, leading to water-dominated atmospheres for strong stellar irradiation and $CO_2$-dominated atmosphere for low stellar irradiation.

This new water-planet model implies that atmospheric composition for water-planets in the HZ should not differ substantially from those of land-ocean-planets, except for an increase in absorbed stellar flux due to the decreased surface reflectivity of water-planets (see S5).

## 3. The Habitable Zone for Water-planets

The "narrow" HZ is defined here classically as the annulus around a star where a rocky planet with a $CO_2/H_2O/N_2$ atmosphere and sufficiently large water content (such as on Earth) can host liquid water on its solid surface (Kasting et al.1993). A conservative estimate of the range of the narrow HZ is derived from atmospheric models by assuming that the planets have a $H_2O$– and $CO_2$–dominated atmosphere with no cloud feedback on the edges of the HZ and determined based on the stellar flux intercepted by the planet (see Kopparapu et al.2013 for details). Cloud feedback widens the limits of the HZ (see e.g. Selsis et al.2007,Zsom et al.2012) but no consistent model exists yet, therefore use the empirical value for the HZ from our own Solar System as the "effective" HZ, defined by the initial solar fluxes received at the orbits of Venus (1Gyr) and Mars (3.5Gyr) when there was no liquid water on their surfaces. In this model it is assumed that the planets are geologically active and that climatic stability is provided by a mechanism in which atmospheric $CO_2$ concentration varies inversely with planetary surface temperature.

The HZ changes only slightly for water-planets compared to land-ocean-planets because at the limits of the HZ the albedo is solely dominated by the atmosphere (see Fig.3 for insight on the effect of the surface albedo on the resulting atmospheric structure). Within the HZ for a given incident stellar flux a water-planet has larger surface temperature due to the low surface reflection of water compared to land.

Cloud coverage influences the atmospheric temperature structure as well as the observability of spectral features (see e.g. Kaltenegger et al.2007, Kaltenegger&Traub 2009, Rauer et al.2010). On Earth cloud fractions is similar over water and land (see e.g. Zsom et al.2012). If sufficient cloud condensation nuclei are available, cloud coverage should be the same as for rocky planets for the same stellar insolation, but it will vary depending on the water-planet's position in the HZ (Rugheimer et al.in prep). The position and pattern of individual cloud layers depend on unknown planetary parameters like rotation rate. We mimic the effects of varying cloud coverage by varying the initial surface albedo in our atmospheric model from the standard value of 0.2 (used by Kasting et al.1993, Kopparapu et al.2013).

The atmospheric structure as well as the resulting HZ limits depend on the density of a planet's atmosphere, shifting the HZ outwards for lower mass and inwards for higher mass planets (see also Kasting et al.1993, Kopparapu et al.2013). Without any further information on the relation of surface pressure to planetary mass, we make the first order assumption here, that if outgassing rates per $m^2$ are held constant on a solid planet, and atmospheric loss rates decrease with planetary mass due to increased gravity, assuming a similar stellar environment, then a more massive planet should have a higher surface pressure (for similar outgassing and atmospheric loss mechanisms). In this paper we scale the surface pressure of the planet to first order with its surface gravity (following e.g. Kaltenegger et al.2011) and explore the effect of different planetary mass in Fig.3.

We use EXO-P (see e.g. Kaltenegger&Sasselov 2010 for details and reference) to model the atmosphere of water-planets. EXO-P consists of a coupled one-dimensional radiative-convective atmosphere code developed for rocky exoplanets based on a 1D-climate, 1D-photochemistry and 1D-radiative transfer model.

## 4. Models for transiting Type1 Water-Planets in the HZ - the Kepler-62 and -69 systems

Kepler-62 (Borucki et al.2013) and Kepler-69 (Barclay et al.2013) are multi-transiting planet systems with individual planets in or close to the HZ (Fig.1 and 2). Kepler-62e, and -62f have radii of 1.61 and $1.41R_E$ with less than 5% errors, orbiting a K2Vstar of 4925±70°K (0.21±0.02$L_0$) at periods of 122.4 and 267.3days, respectively (Borucki et al.2013). They are super-Earth-size planets in the HZ of their host star, receiving 1.2±0.2 and 0.41±0.05 times the solar flux at Earth's orbit ($S_0$). Kepler-69c has a radii of 1.7(+0.34-0.29)$R_E$ orbiting a 5638±168°K Sun-like star and semi-major axis to stellar radius ratio $a/R_{Star}$=148(+20-14) at periods of 242.5days (Barclay et al.2013). Due to the large uncertainty in luminosity of the star Kepler-69c receives 1.91(+0.43-0.56)$S_0$,

The small radii of Kepler-62e, -62f and -69c indicate that the planets should not have retained primordial hydrogen atmospheres at their insolation and estimated ages (Rogers et al.2011,Lopez et al.2013,Lammer et al.2009, Gaidos&Pierrehumbert 2010), making them strong candidates for solid planets, and not Mini-Neptunes. Given the large amount of solid objects already present in the Kepler-62 system on orbits inward of -62e, and following the recent results on the Kepler-11 cohort of 6 low-density planets with tightly packed close-in orbits (Lissauer et al.2013) it strongly suggests that -62e and -62f formed outside the ice line and are therefore our first HZ water-planets. While other possibilities remain open until their actual masses are measured, for the purposes of this paper we'll assume that they are indeed water-planets on low-eccentricity orbits and model both biotic and abiotic atmospheres.

For Kepler-62e we set the $O_2$ mixing ratio to 0.21(biotic), 0.21x$10^{-6}$(abiotic). We set $CO_2$ to 10ppm for Kepler-62e at the inner part of the HZ, which corresponds to the conservative lower limit for C4 photosynthesis (Heath 1969, Pearcy&Ehleringer 1984). For Kepler-62f we calculate what levels of $CO_2$ are needed to maintain liquid water on its surface, what results in 5bar of $CO_2$ for the considered Albedo range. For Kepler-62f we set $O_2$ to 0.21atm for the biotic model because maintaining a constant mixing ratio of 0.21 would yield unrealistically high $O_2$ pressures (about 1 bar). The $CH_4$ flux is set to 1.31x$10^{11}$ (biotic) (Rugheimer et al.2013) and 4x$10^{-9}$ molecules/$cm^2$/s (abiotic) (see e.g. Kasting&Caitling 2003, Segura et al.2007) for both planets.

## 5. Discussion and Conclusions

For Kepler-62e's radius a water-planet's mass would be in the range 2-4$M_E$. Kepler-62f has a smaller radius of 1.41$R_E$, so its mass would be 1.1-2.6$M_E$. The transition from liquid water to the first solid phase of Ice VI occurs at 0.63 to 2.2GPa, depending on the temperature. Beyond 2.2GPa, the solid phase water is Ice VII. The depth of the liquid ocean is thus in the range of 80 to 150km for these two planets. The radius of Kepler-69c $R_p$=1.7(+0.34 -0.23)$R_E$ has very large error bars, therefore the mass range is ill-defined.

Fig.2 shows that the stellar flux values at the limit of the HZ are 1.66 to 0.27$S_\odot$ for the effective HZ and 0.95 to 0.29$S_\odot$ for the narrow HZ, respectively, for Kepler-62. The limits are 1.78 to 0.29$S_\odot$ for the effective HZ and 1.01 to 0.35$S_\odot$ for the narrow HZ for Kepler-69 (based on models by Kopparapu et al.2013). The intercepted flux at the orbit of Kepler-62*e* and -62*f,* make both planets candidates for liquid water on their surface. If the atmosphere of Kepler-62*f* is accumulates several bar of atmospheric $CO_2$ (1.6 to 5bar, depending on the model planetary mass and surface albedo) it could be covered by liquid water on its surface, otherwise it would be ice-covered. This first possibility allows for a warm Type1 water-planet in the inner and outer part of the HZ (Fig.4), the second possibility allows comparing a warm and cold Type1 water-planet in this system. These differences are similar to the variations with temperatures of rocky planet spectra, but water-planets are hotter at a given distance from their star.

Fig.2 shows that Kepler-69*c* lies outside the effective HZ when using to the nominal stellar flux, but the error bar on the star's flux allows the possibility that the planet is within the empirical HZ.

To explore the atmospheric features of transiting water-planet, we concentrate on the Kepler-62 system here, where both planets are within the star's HZ. We run Exo-P models tuned to the Kepler-62 parameters. We also explore the effect of changing cloud parameters by varying the surface albedo from 0.2 to 0.3 mimicing increased water cloud coverage for Kepler-62*e*, and from 0.2 to 0.1 for Kepler-62*f* mimicing decreased water cloud coverage. Fig.3 shows the atmospheric structure of both planets as Type1 water-planets for the minimum mass and explore their changes with surface albedo (nominally 0.2). Interior models (see Zeng&Sasselov 2013) of Kepler-62*e* models give a scaled surface pressure of 0.78 to 1.56 times Earth's and surface temperatures for the low mass case of (A=0.2) 303K and 306K and (A=0.3) 292K and 296K for the abiotic and biotic cases, respectively. Models for the high planetary mass case result in surface temperatures of (A=0.2) 307K and 310K and (A=0.3) 293K and 297K for the abiotic and biotic cases, respectively, showing a slightly higher surface temperature for planets with higher mass and surface pressure.

Atmospheric models for Kepler-62*f* show that 1.6bar (A=0.1) and 5bar (A=0.3) of $CO_2$ is needed to warm the surface temperature above freezing. Models of Kepler-62*f* lead to a scaled surface pressure of 0.56 to 1.32 times Earth's. We add 5bar of $CO_2$ to this pressure, what results in surface temperatures of 288K and 289K(A=0.1) and 280K and 281K(A=0.2) for the abiotic and biotic cases, respectively. For the high mass case of Kepler-62*f*, we obtain surface temperatures of 297K and 298K(A=0.1) and 285K and 287K(A=0.2) for the abiotic and biotic cases, respectively.

Fig.4 uses two model calculations (the lowest surface pressure and gravity consistent with Kepler-62*e* and -62*f*, for biotic conditions) to generate transmission spectra that can inform future instrument sensitivity requirements. This choice allows us to model the maximum observable spectral features among the models, due to the low gravity. Using the highest planetary mass in the interior models would decrease the observable spectral features by a factor of about two.

Fig.4 shows the corresponding synthetic transmission spectra of (top) a water-dominated atmosphere warm water-planet (using model parameters for a light Kepler-62*e*) and (middle) a $CO_2$-dominated atmosphere frozen water-planet (using model parameters for a light

Kepler-62f with only 100PAL $CO_2$), and of a current Earth analog (bottom). The comparison clearly shows stronger $CO_2$ in the transmission spectra of a water-planet on outer edge of the HZ compared to a water-planet the inner edge of the HZ. Therefore relatively low resolution spectra allow such planets to be characterized for telescopes like JWST (see e.g. Kaltenegger&Traub 2009,Deming et al.2009,Rauer et al.2010,Belu et al.2011,vanParis et al.2013) as well as high resolution ground based telescopes like E-ELT (Snellen et al.2013).

We have defined two types of water-planets, Type1, which are true water-planets in their bulk composition and Type2 which are rocky planets with water covering all surface and postulated a new cycling mechanism for $CO_2$ using clathrates. A detailed atmospheric model of the specific planets of Kepler-62e and -62f as Type1 water-planets permits for liquid water on each, if -62f accumulates several bar of $CO_2$ in its atmosphere. We introduce a potential $CO_2$ cycling mechanism on water-planets. The Kepler-62e models show a warm water-planet, between the limits of the narrow HZ and the effective HZ. Fig.4 shows that the transmission spectrum of a water-planet allows an interesting comparison of warm water- versus $CO_2$-dominated as well as warm and cold water-planets in this system. These differences are similar to the variations with temperatures of rocky planet spectra, but water-planets are slightly hotter at a given distance from their star.

Kepler-62e and -62f are the first viable candidates for HZ water-planets which would be composed of mostly solids, consisting of mostly ice (due to the high internal pressure) surrounding a silicate-iron core. The nominal flux intercepted at Kepler-69c is too high for it to be in the effective HZ, but the large error bars on its star's flux allow for this planet close to the inner edge of the empirical HZ.

Water-planets in the HZ are completely novel objects, which do not exist in our own Solar System. The atmospheric models presented here predict detectable features in the spectra of water-planets in the HZ in transit. Therefore we expect that future remote characterization will allow us to distinguish water-planets at different parts of their HZ.

**Acknowledgements:**
The authors Thank J.McDowell for in depth discussions. LK acknowledges support from DFG funding ENP Ka 3142/1-1, DS partial support by NASA NNX09AJ50A (Kepler Mission science team).

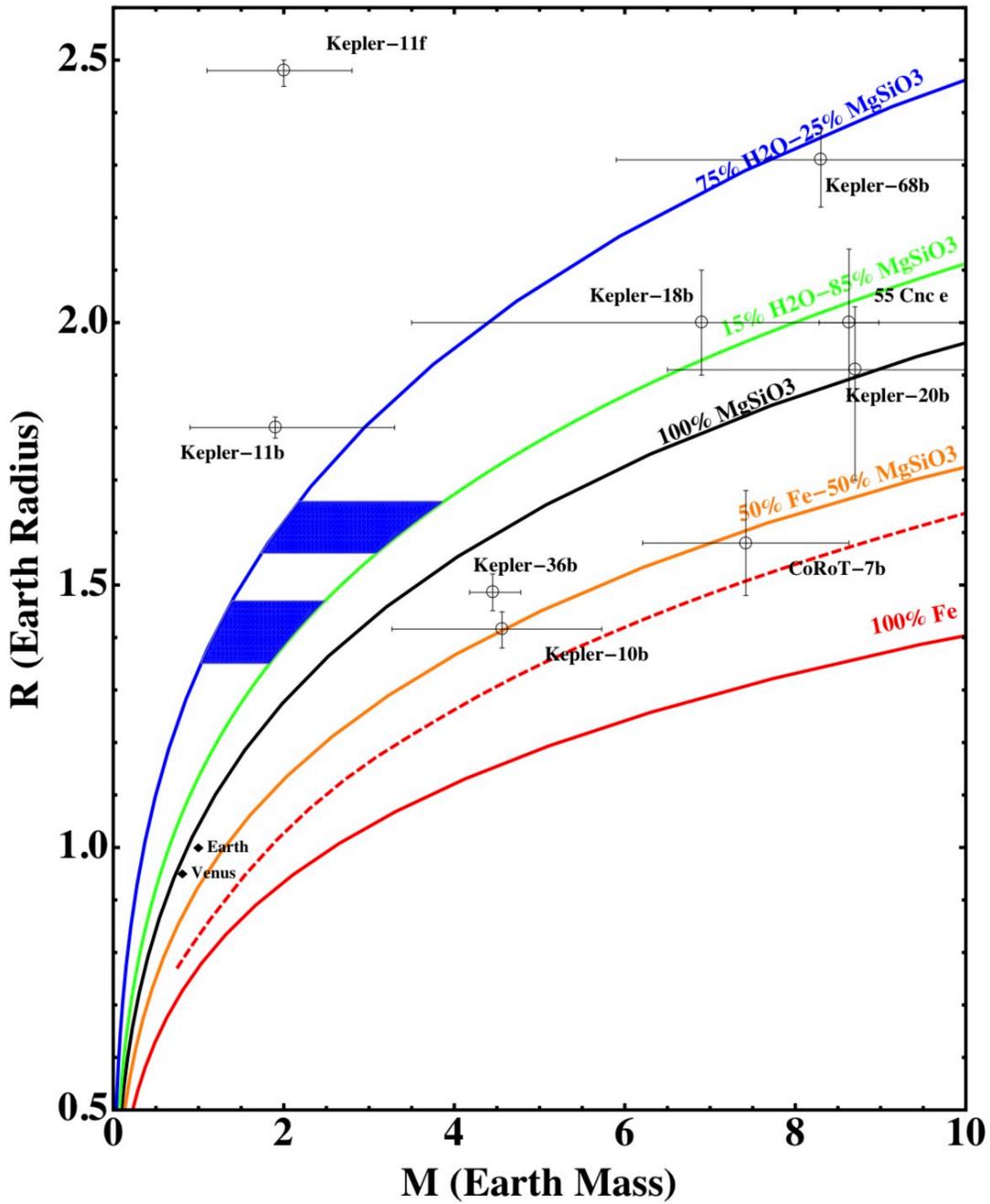

**Fig.1.** Mass-radius plot showing Kepler-62*e* and -62*f* as presumed water-planets (blue lozenges), compared to Venus, Earth, and transiting exoplanets. Kepler-69c ($R_p$=1.7 +0.34-0.23$R_E$) is not shown due to its very large error bars. The curves are theoretical models (Li&Sasselov2013); the dashed line the maximum mantle stripping limit (Marcus et al.2010).

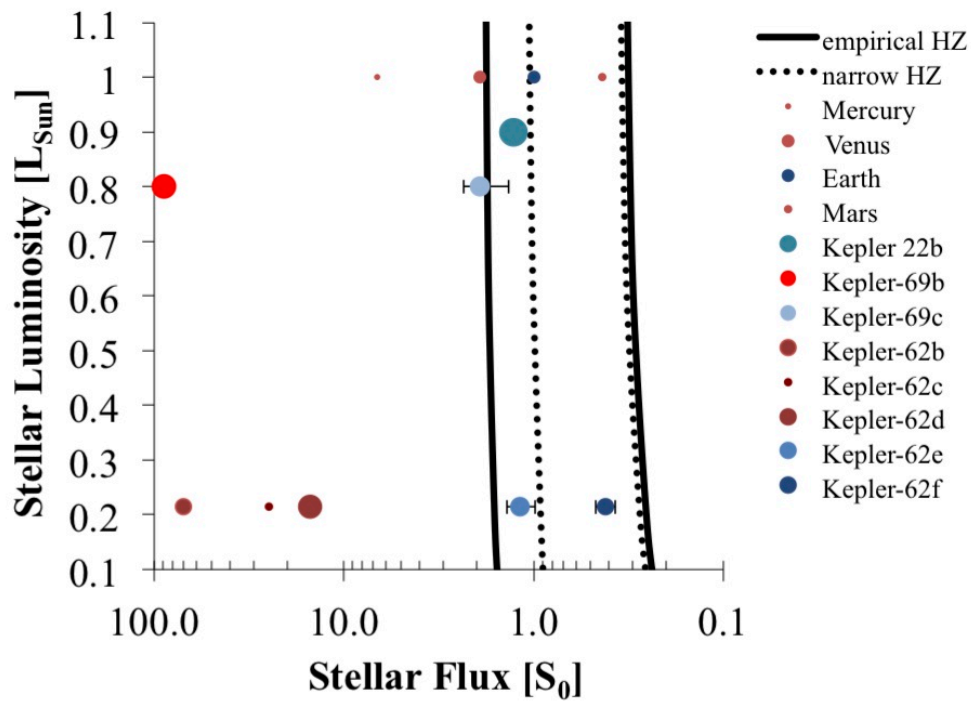

**Fig.2.** Comparison of known transiting exoplanets with measured radii less than 2.5 $R_\oplus$ in the HZ to the Solar System planets. The sizes of the circles indicate the relative sizes of the planets to each other. The dashed and the solid lines indicate the edges of the narrow and effective HZ, respectively.

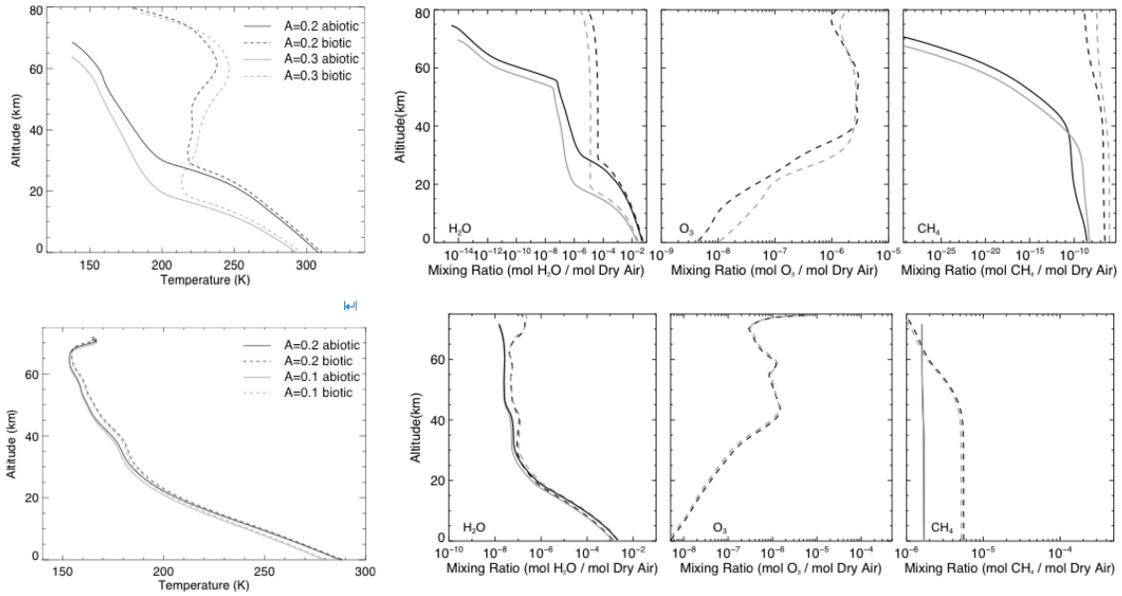

**Fig.3**. Temperature and mixing ratios for $H_2O$, $O_3$ and $CH_4$ for biotic and abiotic atmospheric models for water-planets in the inner(top) and outer (bottom) part of the HZ (using planetary parameters for Kepler-62*e* and −62*f*) for the smallest mass models.

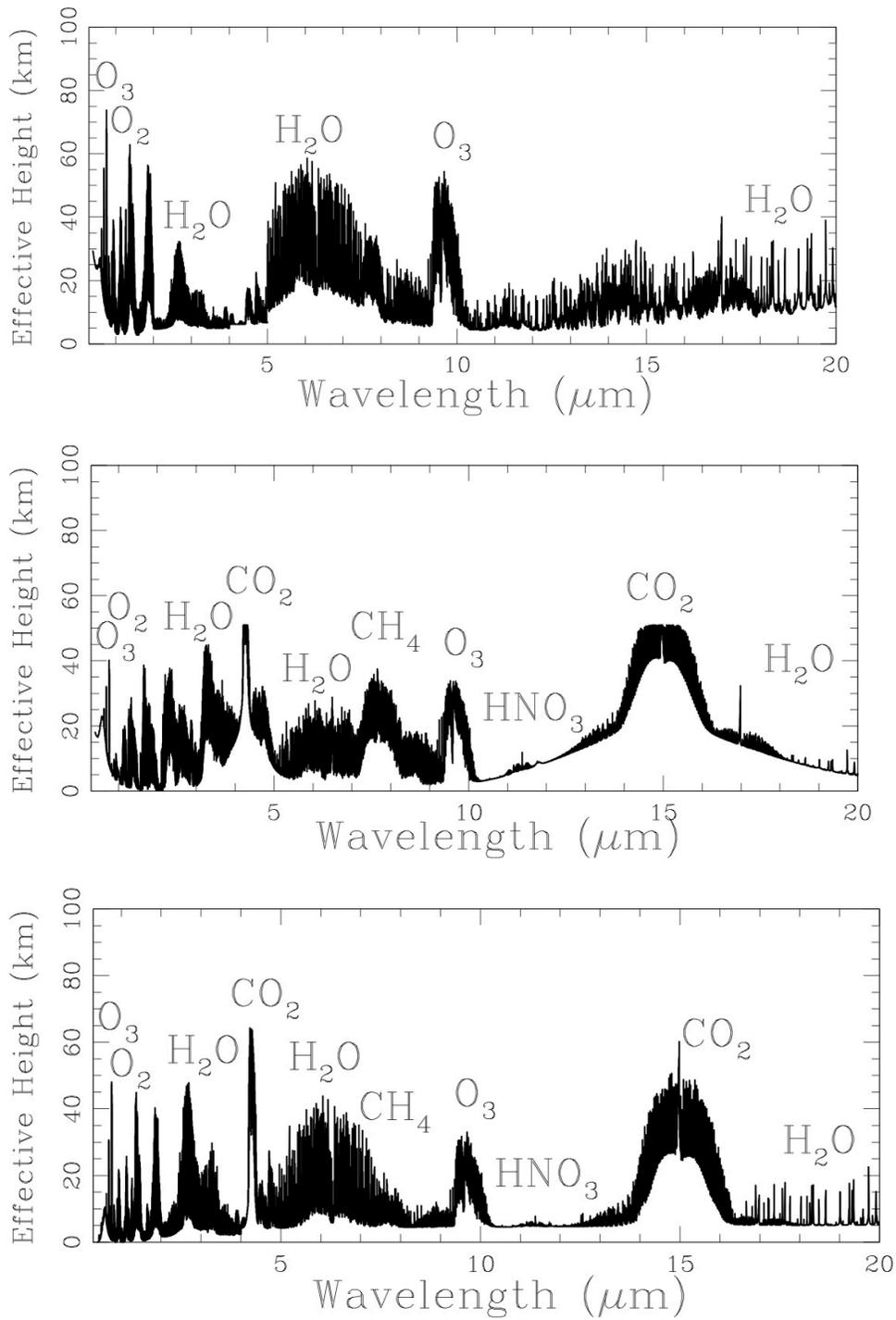

**Fig.4.** Synthetic transmission spectra of atmospheric models for a hot(top) and cold(middle) water-planet, using Kepler-62*e* and -62*f* parameters for the smallest mass models (see text), and the Earth in transit(bottom) for comparison.